\documentclass[journal, letter]{IEEEtran}

\makeatletter
\def\ps@headings{%
\def\@oddhead{\mbox{}\scriptsize\rightmark \hfil \thepage}%
\def\@evenhead{\scriptsize\thepage \hfil \leftmark\mbox{}}%
\def\@oddfoot{}%
\def\@evenfoot{}}
\makeatother
\usepackage{subfigure}
\usepackage{graphicx, amssymb, amsbsy, amsmath}
\usepackage{subfigure}
\usepackage{multirow}
\usepackage{cite,url}
\usepackage{color,soul}
\usepackage[font=footnotesize,labelfont=sf,textfont=sf]{caption}
\usepackage{algorithm}
\usepackage{algorithmicx}
\usepackage{algpseudocode}
\usepackage{setspace}
\graphicspath{{figures/}}
\DeclareGraphicsExtensions{.eps}
\newtheorem{myDef}{\textbf{Definition}}
\newtheorem{theorem}{\textbf{Theorem}}
\def\squareforqed{\hbox{\rlap{$\sqcap$}$\sqcup$}}
\def\qed{\ifmmode\squareforqed\else{\unskip\nobreak\hfil
\penalty50\hskip1em\null\nobreak\hfil\squareforqed
\parfillskip=0pt\finalhyphendemerits=0\endgraf}\fi}

\begin{document}

\title{\huge Follow Me at the Edge: Mobility-Aware Dynamic Service Placement for Mobile Edge Computing}

\author{
	\IEEEauthorblockN{ Tao Ouyang, Zhi Zhou, and Xu Chen*}
    \\School of Data and Computer Science, Sun Yat-sen University, China
    \\ *chenxu35@mail.sysu.edu.cn
    \thanks{Tao Ouyang, Zhi Zhou, and X. Chen,``Follow Me at the Edge: Mobility-Aware Dynamic Service Placement for Mobile Edge Computing,'' accepted by IEEE Journal on Selected Areas in Communications, Aug. 2018.}
	
}
\maketitle

\begin{abstract}
Mobile edge computing is a new computing paradigm, which pushes cloud computing capabilities away from the centralized cloud to the network edge.  However, with the sinking of computing capabilities, the new challenge incurred by user mobility arises: since end-users typically move erratically, the services should be dynamically migrated among multiple edges to maintain the service performance, i.e., user-perceived latency. Tackling this problem is non-trivial since frequent service migration would greatly increase the operational cost. To address this challenge in terms of the performance-cost trade-off, in this paper we study the mobile edge service performance optimization problem under long-term cost budget constraint. To address user mobility which is typically unpredictable, we apply Lyapunov optimization to decompose the long-term optimization problem into a series of real-time optimization problems which do not require a priori knowledge such as user mobility. As the decomposed problem is NP-hard, we first design an approximation algorithm based on Markov approximation to seek a near-optimal solution. To make our solution scalable and amenable to future 5G application scenario with large-scale user devices, we further propose a distributed approximation scheme with greatly reduced time complexity, based on the technique of best response update. Rigorous theoretical analysis and extensive evaluations demonstrate the efficacy of the proposed centralized and distributed schemes.
\end{abstract}
\IEEEpeerreviewmaketitle

\section{Introduction}
\label{sec:intro}
With the explosive growth of mobile devices, the recent years have witnessed an unprecedented shift of user preferences from traditional desktops and laptops to smartphones and other connected devices. Subsequently, more and more new mobile applications, as exemplified by augmented reality and interactive gaming \cite{patel2014mobile}, emerge and catch public attention. In general, these kinds of applications demand intensive computation resources and high energy consumption for real-time processing. However, due to the physical size constraint, the end device can not efficiently support theses applications alone within our expectation. The tension between resource-hungry applications and resource-limited end devices yields a huge challenge for the next generation network development.

A proliferation of powerful and reliable cloud computing, together with widespread fourth/fifth generation (4G/5G) Long Term Evolution (LTE) networks and WiFi access, has brought rich cloud-hosted mobile services \cite{networking2013cisco} to end users. This approach indeed tackles the resource limitation problem of mobile devices. Unfortunately, the long communication latency to the centralized cloud data center (typically hundreds of milliseconds), such as Amazon EC2 and Windows Azure, can far exceed the stringent timeliness requirement (typically tens of milliseconds) of these mission-critical mobile applications. It will significantly deteriorate the user quality of experience. Furthermore, reducing the delay in the wide area network is not tractable.

To satisfy these mission-critical mobile applications that require ultra-low latency, mobile edge computing (MEC) \cite{ar16} \cite{hu2015mobile} has been proposed as an extension of centralized cloud computing, which deploys a cloud computing platform at the edge of radio access network (RAN) \emph{in close proximity to mobile devices and users}. Here an edge is typically a micro-data center or cluster of servers\label{key} that can host cloud applications \cite{chen2018thriftyedge}, attached to a base station (BS) or an access point, and available for use by nearby devices. In the paradigm of MEC, as user workload is served by a nearby edge node rather than the remote cloud, the end-to-end latency is significantly reduced \cite{LiZC18}.
\begin{figure}
	\centering
	\includegraphics[height= 2in,width=3in]{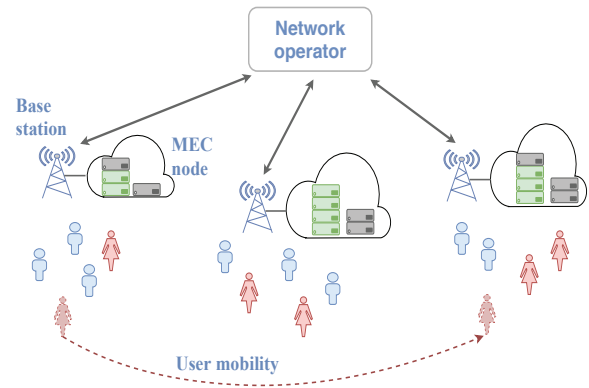}
	\caption{An example of dynamic service placement when a user roams throughout the network in MEC}
	\label{fig-usamap} \vspace{-20pt}
\end{figure}

Although the computation capacity of a mobile user is dramatically augmented by edge cloud, a new challenge arises by unpredicted user mobility in the wireless network. With the presence of user mobility \cite{Xenakis2014}, enhancing low-latency and smoothing user experience are far more than simply pushing the cloud capabilities to the network edge. To guarantee service continuity when users travel across different edges, an efficient mobility management scheme should be employed in the network edge. An emerging technique, software-defined network (SDN) \cite{afolabi2018network} is proposed to provide seamless and transparent mobility support to users. In the SDN based fog computing architecture \cite{bi2018mobility}, the routing logic and intelligent management logic are deployed on the SDN controller, which dramatically simplifies network operation and management. Let us consider a practical scenario as shown in Fig. 1, when a mobile user is within the geographical coverage of the left MEC node, it is clear that if we want to minimize the user-perceived latency, the user should be served by the nearest edge, i.e., the left one. Considering the user mobility and assuming that after a while, the aforementioned mobile user moves to the coverage of the right MEC node. Then, if the service profile of this user is still placed at the left MEC node to serve this user, his perceived latency would greatly deteriorate due to the extended network distance. This example demonstrates that, to optimize the user experience of MEC, the service profiles of mobile users should be dynamically re-placed among edges to \emph{follow the mobility} of users.

However, the dynamic service profile placement problem is non-trivial. On one hand, the user-perceived latency is jointly determined by the communication delay and computing delay \cite{chen2017exploiting} \cite{chen2018efficient}. Therefore, if the service profile of each user is placed aggressively at the nearest MEC node, then some MEC nodes could be overloaded, leading to increased computing latency. On the other hand, following the user mobility requires frequent service migration among multiple MEC nodes. In return, such frequent service migration incurs additional operational cost such as usage of the expensive wide-area-network (WAN) bandwidth and system energy consumption \cite{Cong2016}. As a result, an effective dynamic service placement strategy should carefully (1) cooperate the communication delay and computing delay to minimize the user-perceived latency, and (2) navigate the \emph{performance-cost trade-off} in a cost-efficient manner.

Following the above two guidances on dynamic service placement for MEC, in this paper, we propose a \emph{mobility-aware dynamic service placement framework for cost-efficient MEC}. In particular, to strike a nice balance between the service performance and the operational cost incurred by cross-edge service migration, we propose to minimize the user-perceived latency over the long run, under the constraint of a long-term migration cost budget which is pre-defined monthly or yearly by the network operator in practice. By applying Lyapunov optimization technique to the formulated stochastic optimization problem, our framework can effectively incorporate the long-term migration cost budget into real-time optimizations, and make online decisions on dynamic service placement, \emph{without requiring any a priori future information (e.g., user mobility)}. To address the challenge of the NP-hardness of the resulted real-time optimizations, we further design centralized and distributed approximation schemes to seek near-optimal solutions respectively. Both rigorous theoretical analysis and extensive trace-driven evaluations demonstrate the cost-efficiency of the proposed mobility-aware online service placement framework. 

The rest of this paper is organized as follows. 
Section~\ref{related} reviews related work. The system model and problem formulation are presented in Section~\ref{basic}. Section~\ref{simulation} proposes two online service placement algorithms under the centralized and decentralized mechanisms to seek near-optimal strategies respectively. Section~\ref{verification} presents the theoretical analysis of the proposed framework. Performance evaluation is carried out in Section~\ref{OCA}. Section~\ref{sec:con} concludes this paper.

\section{Related Work}
\label{related}

Service placement is not a new topic, as it has been extensively studied in the paradigm of cloud computing. Specifically, the goal of service placement in cloud computing can be categorized into: (1) consolidating the services to a smaller set of physical servers to improve the resource utilization and reduce the operational cost \cite{JVM12}, (2) placing the services to a set of heterogeneous nodes to leverage the heterogeneities on energy efficiency or cost efficiency \cite{Xu2013Managing}, and (3) placing the services to different nodes to perform network load balancing \cite{VN13}. However, as we have discussed in Section~\ref{sec:intro}, the goal of service placement in MEC is to follow the user mobility and thus to reduce the user-perceived latency adaptively. 

A key challenge towards efficient service placement in MEC is to follow the mobility of users and devices. In addressing this challenge, some work is based on the assumption of perfect predictability on future information. For example, the authors in \cite{MP16} tackle the trade-off between the execution overhead and latency, with a mobility-based prediction scheme which estimates the data transfer throughput, handoff time and VM migration management in advance. Moreover, the recent work \cite{DS17} further studies how to place service by predicting the future cost incurred by data transmission, processing and service migration. \cite{aissioui2018enabling} proposes a FMeC-based framework in an automated driving use case, which captures the trade-off between reducing service migration cost and maintaining the end-to-end QoS based on the vehicle mobility pattern update analysis. But this work does not consider the load balancing among multiple edge servers in the multi-user case. Unfortunately, the future information such as user mobility is extremely challenging to accurately predict in realistic environments.

In response to the challenge that user mobility may not be readily predictable in practice, another stream of recent work resorts to a milder assumption that the user mobility follows a Markovian process, and then applies the technique of Markov Decision Process (MDP). Specifically, a preliminary research in \cite{tt13} explores how service migration impacts the perceived latency of mobile users, via utilizing Markov chains to analyze whether to migrate services or not. Both \cite{ksentini2014markov} and \cite{SW2014} try to determine an optimal threshold decision policy on service migration based on MDP. Further, \cite{T16} extends service migration decision algorithm \cite{ksentini2014markov} to capture 2D
mobility scenarios. In \cite{SW15}, the optimal service migration strategy is devised by formulating the service placement problem as a sequential decision-making problem. In comparison, our online service placement strategy does not make any assumption on the user mobility, yet can achieve a  performance that can be arbitrarily close to the offline optimum. Moreover, all these works do not consider the practical operational cost constraint for dynamic service placement.

Without requiring the future user mobility as a priori knowledge, \cite{kiani2017toward} designs a two-time scale approach to maximize the profit of a service provider, while satisfying users' QoS by allocating computing and communications resources in hierarchical mobile edge computing. A pronounced difference is that our work considers a long-term cost budget constraint, where our algorithm can be continuously adjusted to accommodate system dynamics. A closely related
work \cite{EMM17} proposes an energy-aware mobility management scheme to minimize the total delay (including both communication and computation delay) under the long-term energy consumption constraint. However, it is worth noting that our work substantially differs from and complements to \cite{EMM17} in at least the following three aspects: (1) the study in \cite{EMM17} only considers a single-user service placement scenario, while we consider a more practice-relevant multi-user case. (2) We consider the efficient allocation of the limited edge resource to multiple users, and thus to coordinate computation and communication delay to minimize the total latency. (3) To avoid excessive operational cost incurred by frequent service migration, we navigate the performance-cost tradeoff in a cost-efficient manner. 

This work significantly extends the preliminary work \cite{ouyangfollow}. To improve the service performance in a large scale and ultra dense network, we propose a distributed service placement scheme with faster convergence rate in this work. Rather than optimizing the service performance based on Markov approximation in a cooperative manner, the distributed scheme minimizes the cost in a non-cooperative manner. To evaluate the efficiency of distributed algorithms, we introduce a dynamic placement policy with mobility pattern update analysis \cite{aissioui2018enabling} and more detailed comparison between two proposed algorithm (such as the running time of decision making with different dense works) in the experiment.

\begin{table}
	
	\renewcommand\arraystretch{1.2}
	\centering
	\caption{Key notations in our model}
	\setlength{\arrayrulewidth}{0.6pt}
	\begin{tabular}{|c||p{2.7in}|}
		\hline 
		Notation & \multicolumn{1}{c}{Definition} \vline \\
		\hline
		\hline
		\multirow{2}*{$x_i^k(t)$} & Whether the service profile of user k is placed at MEC node i (=1) or not (=0)\\
		\hline
		\multirow{2}*{$R^k(t)$}&  The amount of computation capacity required by user $k$\\
		\hline
		$N_i(t)$ & The number of services served by MEC node $i$\\
		\hline
		\multirow{2}*{$F_i$} & The maximum computing capacity of MEC node $i$\\
		\hline
		$D^k(t)$ & The computing delay for user $k$ \\
		\hline
		\multirow{2}*{$H_i^k(t)$} & The communication delay when the service profile of user k is placed on MEC node $i$\\
		\hline
		$L^k(t)$ & The communication delay for user $k$\\
		\hline
		$T^k(t)$ & The total perceived latency for user $k$\\
		\hline
		\multirow{2}*{$E_{ji}^k(t)$} & The cost of migrating service of user $k$ from source MEC node $j$ to destination MEC node $i$\\ 
		\hline
		$E(t)$ & The total migration cost for all users\\
		\hline
		$E_{avg}$ & The long-term time-averaged cost budget\\ 
		\hline
		
		$V$ & Lyapunov control parameter\\
		\hline
		$\beta$ & Markov approximation control parameter\\
		\hline
		$\mu$ & The approximation ratio of distributed scheme\\
		\hline
	\end{tabular}
	\label{tab1}

\end{table}
\section{System Model and Problem Formulation}
\label{basic}
As shown in Fig.1, we consider a network operator running a set ${\mathcal{M}} = \{1,2,...,M\}$ of MEC nodes to serve a set ${\mathcal{N}} = \{1, 2,..., N\}$ of mobile users. Each MEC node is attached to a local base station or wireless access point, via high-speed local-area network (LAN). Inspired by the recent work \cite{Urgaonkar2015} \cite{Mach2017} on resource allocation for MEC, in this paper we adopt a device-oriented service model for MEC, rather than the traditional application-oriented service model for cloud computing \cite{zhi2013saas}. Specifically, the service profile and environment for the applications run on each mobile device (rather than each application) is assigned to a dedicated virtual machine \cite{Chaufournier2017} or container \cite{William}. To better capture the user mobility, the system is assumed to operate in a slotted structure and its timeline is discretized into time frames $t \in \mathcal{T} = \{0,1,2..., T\}$. At all discrete time slots, each mobile user sends a service request to the local MEC node, then the network operator (i.e., SDN controller) will gather all request information and determine the optimal MEC node to serve corresponding user based on the current global system information. Table \ref{tab1} summarizes the key
parameter notations in our paper.
\subsection{Service Placement Model}
To maintain satisfactory Quality-of-Service (QoS), i.e., low service latency for mobile users which typically move erratically, the service profile of each user should be dynamically migrated across multiple edges to follow the user mobility. Here we take a binary indicator $x_i^{k}(t)$ to denote the dynamic service placement decision variable. Let $x_i^{k}(t)=1$ if the service profile of user $k \in \mathcal{N}$ is placed at the MEC node $i \in \mathcal{M}$ at time slot $t$, and $x_i^{k}(t)=0$ else. Note that at a given time slot, since each user is served by one and only one MEC node, we have the following constraints for service placement decision $x_i^{k}(t)$:
\begin{equation}
\sum_{i=1}^M x_i^{k}(t) = 1, \quad \forall k, t.
\end{equation}
\begin{equation}
x_i^{k}(t) \in \{0, 1\}, \quad \forall i, k, t.
\end{equation}
Based on the above defined service placement decision, we are now ready to formulate the user-perceived latency which is determined by the service placement.

\subsection{QoS Model}
In the paradigm of MEC, the QoS, i.e., user-perceived latency is jointly determined by the computing delay and communication delay. 

\textbf{Computing delay:} at each MEC node, multiple mobile users will simultaneously share the computing resource to process their applications. 
However, when confronted with service request surge, the small-scale MEC node may not guarantee to provide satisfactory service for all served users. A more efficient load balancing across multiple MEC nodes can be achieved by dynamic service placement. In this paper, we use $R^k(t)$ to denote the amount of computing capacity (in terms of CPU cycles) required by the service request of user $k$ at time slot $t$.
Taking video stream analytics as an instance \cite{EMM17}, the amount of required computation capacity is determined by the input data size of the video and the corresponding computation intensity of the analytic task. 
We consider an equal resource allocation case, i.e., each user evenly shares computing resource of the serving MEC node\footnote{Other resource allocation models, such as weighted resource allocation, are also applicable to Markov approximation based scheme.}. 
Then, the computing delay for mobile user $k$ at time slot $t$ is given by $D^{k}(t) = \sum_{i = 1}^{M}x_i^k(t) R^k(t)N_i(t)/F_i,$ where $N_i(t)$ is the number of users served by MEC node $i$ during the time slot $t$, which follows $N_i(t) = \sum_{k = 1}^{N}x_i^{k}(t).$ Moreover, $F_i$ represents the maximum computing capacity (in CPU cycles per second) of MEC node $i$.

\textbf{Communication delay:}  in MEC, the communication delay between a mobile device and the MEC node generally contains the network propagation delay and the data transmission delay. In particular, when a data packet passes through the intermediate network devices along the targeted path between service-served MEC node and local connected MEC node, the network propagation delay is majorly determined by the network distance (i.e., the hop count), such as in \cite{T16}. While the data transmission delay is jointly determined by the amount of data transferred $d^k$ and the link bandwidth $b_i^k$. Then the delay can be denoted as $\gamma\frac{d^k}{b_i^k}$, where $\gamma$ is a positive coefficient. Since the network condition (i.e., hop distance and bandwidth) and data transmission information in current time slot are available from the system-level perspective, we can extend the above cases into a general model, which we do not impose structural assumption on.  
Given the service request information as well as the current location of user $k$, the communication delay to MEC node $i$ can be characterized by a general model $H^k_i(t)$. When considering the service placement decision $x_i^{k}(t)$, the communication latency experienced by user $k$ can be further expressed as $L^{k}(t) = \sum_{i = 1}^{M}x_i^k(t)H^k_i(t).$

By combining the computing delay $D^{k}(t)$ and communication delay $L^{k}(t)$, we denote the total latency experienced by user $k$ at time $t$ as
\begin{equation}
T^k(t) = D^k(t) + L^k(t).
\end{equation}
\vspace{-30pt}

\subsection{Migration Cost Model}
While dynamic service placement empowers satisfactory QoS by migrating service profiles among edges to follow the user mobility, it is worth noting that cross-edge service migration would incur additional operational cost. Specifically, when transferring the service profile of each user across edges, enormous usage of the scarce and expensive wide-area-network (WAN) bandwidth would be caused. In additional, cross-edge transferring also increases the energy consumption of network devices such as routers and switches. To model the operational cost incurred by cross-edge service migration, we use $E_{ji}^{k}(t)$ to denote the cost of migrating the service profile of user $k$ from source MEC node $j$ to destination MEC node $i$. Without loss of generality, we assume that $E_{ji}^{k}(t)=0, \forall j=i$. Then, given the service placement decision $x_i^k(t-1)$ at time slot $t-1$, and $x_i^k(t)$ at time slot $t$, the service migration cost of user $k$ at time slot $t$ can be computed by $\sum_{i=1}^{M}\sum_{j=1}^{M}x_j^{k}(t-1)x_i^{k}(t)E_{ji}^{k}(t)$. Considering all the $N$ users, the total service migration cost at time slot $t$ can be further denoted as
\begin{equation}
E(t) = \sum_{k =1}^{N}\sum_{i=1}^{M}\sum_{j=1}^{M}x_j^{k}(t-1)x_i^{k}(t)E_{ji}^{k}(t). \nonumber
\end{equation}
With the presence of user mobility, it is intuitive that to ensure a desirable level of QoS, the service profile should be actively migrated to follow the user mobility. However, frequent migration would incur excessive operational cost in return. Then, a natural question is how to navigate such a performance-cost trade-off in a cost-efficient manner.

\subsection{Navigating the Performance-Cost Trade-off}
To optimize multiple conflicting objectives in a balanced manner, the most commonly adopted approach is to assign different weights to those conflicting objectives and then optimize the weighted sum of them. Unfortunately, in our problem, how to properly defining the weights of performance and cost in a realistic environment is not straightforward. In response, considering the fact that network providers generally operate within a long-term (e.g., yearly) cost budget, we propose to optimize the long-term performance under the predefined long-term cost budget. Specifically, we introduce $E_{avg}$ to denote the long-term time-averaged cost budget over a time span of $T$ time slots, which satisfies:
\begin{equation}
\label{long-termConstraint}
\lim_{T \to \infty} \frac{1}{T}\sum_{t = 1}^{T} E(t)\le E_{avg}.
\end{equation}
Then, our problem of minimizing the long-term time-average service latency under the constraints of long-term cost budget can be formulated as the following stochastic optimization:
\begin{eqnarray}
\label{P1}
\begin{aligned}
{\mathcal{P}1}: \quad &\min_{c(t)} \quad\frac{1}{T}\lim_{T \to \infty} \sum_{t = 1}^{T}\sum_{k = 1}^{N}T^k(t) \\
& \rm{s.t.} \quad (1)-(4). \\
\end{aligned}
\end{eqnarray}
In general, the derivation of the optimal long-term policy ${\mathcal{P}1}$ is not a one-shot operation but needs to be continuously adjusted to accommodate system dynamics such as erratic user mobility and requested service pattern. This is because predicting accurate user behavior (mobility and requested service pattern) and network condition over a long run is extremely hard. Moreover, even though the long-term service placement optimization has been decomposed into the real-time decoupling problem, preventing frequent service migration with the long-term migration cost constraint is non-trivial. In the current literature, some approaches have been proposed to handle this problem. For example, in \cite{DS17}, by finding an optimal look-ahead window size, the long-term optimization problem can be approximately discretized into a series of equivalent shortest-path problems. However, the near-future information cannot be predicted accurately for dynamic mobile wireless network. Fortunately, in the queuing theory \cite{MJ2010}, the long-term migration budget constraint (\ref{long-termConstraint}) in this optimization problem can be regarded as the queue stability control, i.e., the time-averaged migration $\lim_{T \to \infty} \frac{1}{T}\sum_{t = 1}^{T} E(t)$ is beneath the long-term budget $E_{avg}$. Moreover, Lyapunov optimization technique provides an efficient approach to decouple the long-term problems. It does not require any a priori system information while maintaining the queue stability in an online way. Hence, we propose an online algorithm that transforms the original problem into a series of real-time minimization problems.

\section{Online Service Placement Algorithm}
\label{simulation}
In this section, we describe a novel framework that makes online service placement decisions. To solve the $\pmb{\mathcal{P}1}$, we first convert the original problem to a queue stability control problem based on Lyapunov optimization.
\subsection{Problem Transformation via Lyapunov Optimization}
\subsubsection{The construction of virtual queue for long-term service placement cost}
Due to the dynamic and stochastic property of the system (e.g., time-varying and uncertainly user mobility and request arrival process), a prime challenge of $\pmb{\mathcal{P}1}$ is to navigate the performance-cost trade-off in a cost-efficient manner without global information over the long run. A key idea of Lyapunov optimization is to strike a desirable balance between current perceived latency and migration cost while maintaining the cost queue stable by introducing a virtual queue for the long-term budget. First, we define a virtual queue as a historical measurement of the exceeded migration cost and assume that initial queue backlog is 0 (i.e., $Q(0) = 0$).
\begin{eqnarray}
\label{expressionOfQueue}
Q(t+1) = max[Q(t) + E(t) - E_{avg}, 0],
\label{virtual queue}
\end{eqnarray}
where $Q(t)$ is the queue length at time slot $t$, which represents the exceeded cost of executed service migration by the end of time slot $t$.

Intuitively, the value of $Q(t)$ can be regarded as an evaluation criteria to assess the migration cost condition. A large value of $Q(t)$ implies the cost has far exceeded the long-term budget $E_{avg}$ since carrying out online service placement algorithm. In order to guarantee that the time-averaged service migration cost is lower than budget $E_{avg}$, i.e., inequality (\ref{long-termConstraint}) holds, the virtual queue $Q(t)$ must be stable, i.e., $\lim_{T \to \infty}  \mathbb{E}\{Q(T)\}/T=0 $. Furthermore, by total summing the inequality $Q(t+1) \geq Q(t) + E(t)-E_{avg}$ derived from equation (\ref{expressionOfQueue}) and rearranging it, we can gain:
\begin{displaymath}
	\frac{Q(T) - Q(0)}{T} + E_{avg} \geq \frac{1}{T}\sum_{t=0}^{T-1}E(t).
\end{displaymath}
For $Q(0) = 0 $, we can take expectations of the above inequality and have 
\begin{displaymath}
\lim_{T \rightarrow \infty}\frac{\mathbb{E}\{Q(T)\}}{T} + E_{avg} \geq \frac{1}{T}\sum_{t=0}^{T-1}\mathbb{E}\{E(t)\}.
\end{displaymath}  
Hence, the stability of the virtual queue can ensure that the time-averaged migration cost does not exceed the budget.
\subsubsection{Queue stability}
To stabilize the virtual queue, we first define a quadratic Lyapunov function and Lyapunov drift function \cite{neely2010stochastic} respectively as follows:
\begin{eqnarray}
L(\Theta(t)) \triangleq \frac{1}{2}Q(t)^{2}.
\end{eqnarray}
This represents a scalar measure of cost queue congestion. For instance, a small value of $L(\Theta(t))$ implies the queue backlog is small. Thus, if a policy consistently pushes the quadratic Lyapunov function towards a bounded level, it implies that the virtual queue is stable.

To remain the virtual queue stable, we introduce the \emph{one-step conditional Lyapunov drift} to push the quadratic Lyapunov function towards a lower congestion region:
\begin{eqnarray}
\Delta(\Theta(t)) \triangleq \mathbb{E}\Big[L(\Theta(t + 1)) - L(\Theta(t)) | 	\Theta(t)\Big].
\end{eqnarray}
The drift $\Delta(\Theta(t))$ denotes the migration cost queue change in the Lyapunov function over a one-time slot. It generates an important term that includes a product of queue backlog and migration cost,
which helps the algorithm adjust to accommodate the system dynamics \cite{neely2010stochastic}.
\subsubsection{Joint Lyapunov drift and user-perceived latency minimization}	
After constructing the virtual cost queue, the original problem has been decomposed into a series of real-time optimization problems. Our goal is to find a current placement policy to coordinate the perceived latency and migration cost. By incorporating queue stability into delay performance, we define a \emph{Lyapunov drift-plus-penalty} function to solve the real-time problem.
\begin{eqnarray}
\Delta(\Theta(t)) + V\sum_{k = 1}^{N}T^k(t),
\end{eqnarray}
where $V$ is a non-negative control parameter that adjusts the trade-off between delay performance and migration cost queue backlogs. It shows the attention on the delay performance compared to migration cost budget. Moreover, the following lemma provides the performance guarantee of the drift-plus-penalty function.
\newtheorem{lemma}{\textbf{Lemma}}
\begin{lemma}
\label{lemma-drift}
 For all possible values of $\Theta(t)$ by using any placement schedule over all time slots, the following statement holds:
\begin{eqnarray}
\begin{aligned}
\Delta(\Theta(t)) + V\sum_{k = 1}^{N}T^k(t) &\le  
B+ \sum_{k = 1}^{N}V\mathbb{E}\Big[T^k(t)|\Theta(t)\Big] \\  
&+  Q(t)\mathbb{E}\Big[E(t) - E_{avg}|\Theta(t)\Big],
\end{aligned}
\end{eqnarray}
where $B=\frac{1}{2}(E_{avg}^{2}+E^{2}_{max})$ is a constant value for all time slots, and $E_{max}=\max_{t \in \mathcal{T}}E(t)$.
\end{lemma}

The detailed proof is given in the technique report \cite{proof}. Based on the \textit{Lemma 1}, the \emph{drift-plus-penalty} function has a supremum bound at every time slot $t$.
\subsection{Online Service Placement Algorithm}
In this section, we convert the problem $\pmb{\mathcal{P}1}$ to a series of real-time drift-plus-penalty supremum bound minimizations. While the \emph{drift-plus-penalty} expression involves the $max[*]$ term in equation (\ref{expressionOfQueue}), which complicates reaching solution to placement issue. Following the lemma 1, we observe that minimizing the right side of inequality (10) can approximate the supremum bound closely, which is equivalent to minimizing the \emph{drift-plus-penalty}. Therefore, based on the aforementioned parameters definition, we rearrange it for a concise form and obtain an optimal service placement policy $c^*(t)$. 
\begin{small}
\begin{eqnarray}
\begin{aligned}
&\sum_{k = 1}^{N}V\mathbb{E}\Big[T^k(t)|\Theta(t)\Big]+Q(t)\mathbb{E}\Big[E(t) - E_{avg}|\Theta(t)\Big]\\ &\le \sum_{k = 1}^{N}\sum_{i=1}^{M}x_i^{k}(t)\Big(\frac{VR^k(t)\sum_{k = 1}^{N}x_i^{k}(t)}{F_i}+V H^k_i(t) + \rho_i^k(t)\Big), \\
\end{aligned}
\end{eqnarray}
\end{small}
where $\rho_i^k(t) = \sum_{j=1}^{M}x_j^{k}(t-1)Q(t)E_{ji}^k$ is a constant at every time slot $t$, which does not affect the placement decision-making. Thus, the major part of our online service placement algorithm is to solving following $\pmb{\mathcal{P}2}$ to minimize the real-time supremum bound for the \emph{drift-plus-penalty} function.
\begin{small}
\begin{equation}
\label{P2}
\begin{aligned}
\pmb{\mathcal{P}2}: &\min_{c(t)} \sum_{k = 1}^{N}\sum_{i=1}^{M}x_i^{k}(t)\Big(\frac{VR^k(t)\sum_{k = 1}^{N}x_i^{k}(t)}{F_i} + V H^k_i(t) + \rho_i^k(t)\Big) \\
& \rm{s.t.} \quad (1)-(4). \\
\end{aligned}
\end{equation}
\end{small}
For simplify the formulation, we use $U(\textit{\textbf{c},{t}})$ to replace the objective function of problem  $\pmb{\mathcal{P}2}$, where $\textit{\textbf{c}}$ is feasible service placement policy. In Algorithm 1, we describe the implementation of the online service placement algorithm. In each time slot $t$, a close-to-optimal service placement schedule can be obtained when solving $\pmb{\mathcal{P}2}$, and the migration cost virtual queue will be updated subsequently for next time slot calculation.  

\begin{algorithm}[H]
	\caption{Online Service Placement Algorithm} 
	\begin{algorithmic}[1]
		\State \textbf{Initialization}: We set the cost queue backlog $Q(0) = 0$ at beginning.   
		\State \textbf{End initialization}                             
		\For{each time slot $t = 1,2,...,\infty$}
		\State \textbf{Solve} the problem $\pmb{\mathcal{P}2}$: $\textbf{c}^*(t) = arg\min(\ref{P2})$.
		\State \textbf{Update} the virtual queue: run (\ref{expressionOfQueue}) based on $\textbf{c}^*(t)$.
		\EndFor  
	\end{algorithmic}    
\end{algorithm}	
\vspace{-5pt}
Unfortunately, this real-time optimization problem is NP-hard in general \cite{blumrosen2006welfare}, due to its combinatorial nature. To address this challenge, we apply Markov approximation \cite{Chen2013} to obtain a near-optimal solution for this real-time problem.

\subsection{Markov Approximation Method}
In this subsection, we design a centralized service placement optimization that can obtain the minimum solution approximatively. The problem $\pmb{\mathcal{P}2}$ is a combinatorial optimization of finding the optimal service placement policy, we leverage the idea of Markov approximation in \cite{Chen2013} to optimize the policy. To proceed, then we can convert the problem $\pmb{\mathcal{P}2}$ to the following equivalent problem:
\begin{eqnarray}
\label{P3}
\begin{aligned}
& \min \quad \sum_{\textbf{c}\in c(t)}q_\textbf{c}(t)U(\textit{\textbf{c},{t}}) \\
& s.t. \quad \sum_{\textbf{c}\in c(t)}q_\textbf{c}(t) = 1, \forall t \in \mathcal{T},\\
\end{aligned}
\end{eqnarray}
where $q_\textbf{c}(t)$ is a decision variable, which means the probability of the placement policy $\textbf{c}$ is adopted at current time slot $t$; $c(t)$ is the collection of all feasible placement policies. Obviously, the optimal solution to problem (\ref{P3}) is to choose the minimum cost placement policy with probability one. The problem can be approximately treated as the following \emph{convex log-sum-exp} problem \cite{Chen2013}. 
\begin{eqnarray}
\begin{aligned}
\label{P4}
&\min \quad \sum_{\textbf{c}\in c(t)}q_\textbf{c}(t)U(\textit{\textbf{c},{t}}) + \frac{1}{\beta}\sum_{\textbf{c}\in c(t)}q_\textbf{c}(t)\log q_\textbf{c}(t)\\
& s.t. \quad \sum_{\textbf{c}\in c(t)}q_\textbf{c}(t) = 1,\forall t \in \mathcal{T}, \\
\end{aligned}
\end{eqnarray}
where $\beta$ is a positive constant that charges the approximation ratio of the entropy term. When $\beta \rightarrow \infty$, the problem (\ref{P4}) becomes the original problem (\ref{P3}). If handling the problem with the Karush-Kuhn-Tucker (KKT) \cite{Boyd2010}, we can obtain the optimal solution to problem (\ref{P4})
\begin{eqnarray}
\label{optimalPolicy}
q_\textbf{c}^*(t) = \frac{exp\big(-\beta U(\textit{\textbf{c},{t}})\big)}{\sum_{\textbf{c'} \in c(t)}exp\big(-\beta U(\textit{\textbf{c'},{t}})\big)}, \forall \textbf{c} \in c(t), \forall t \in \mathcal{T}.
\end{eqnarray}

\begin{algorithm}[H]
	\caption{Markov Approximation based Placement Policy Search} 
	\begin{algorithmic}[1]
		\State \textbf{Initialization:} Initialize the service placement policy \textbf{c} as randomly assigning a MEC node for each service.   
		\State \textbf{End initialization}
		\Loop{  for each service placement update iteration}
		\State \textbf{Choose} a service $k$ randomly and carry out the following operations:
		\State \quad \textbf{Calculate} the bound $U(\textit{\textbf{c$'$},{t}})$ for any other feasible service placement policy.
		\State \quad \textbf{Select} a placement policy acc. to (\ref{transfer_prob}) probabilistically. 
		\State \quad \textbf{Update} the service placement policy by placing the service to the new MEC node.		
		\State \textbf{Record} the placement policy $\textbf{c*}$ with the smallest $U(\textit{\textbf{c*},{t}})$, found up to now. 
		\EndLoop
	\end{algorithmic}    
\end{algorithm}
According to the probability $q_\textbf{c}^*(t)$, we can gain the current optimal policy. Then, we design a service placement algorithm that constantly updates the placement policy $\textit{c}$ to form a discrete-time Markov chain \cite{Chen2013}. Once the Markov chain achieves to the stationary distribution as shown in (\ref{stationaryDistribution}), the optimal placement profile which minimizes the real-time supremum bound in (\ref{P3}) can be derived by setting parameter $\beta$ as large as possible. In this algorithm, the Markov chain is irreducible, which traverses all feasible states under different placement policies. Besides, designing a desired time-reversible Markov chains needs to hold the following balance equation:
\begin{equation}
\label{stationaryDistribution}
	q_{\textbf{c}}^*(t) q_{\textbf{c},\textbf{c}'}(t) = q_{\textbf{c}'}^*(t) q_{\textbf{c}',\textbf{c}}(t), \forall\textbf{c},\textbf{c}' \in c(t),	\forall t \in \mathcal{T},
\end{equation}   
where $q_{\textbf{c},\textbf{c}'}(t)$ is the probability of the placement policy update from $\textbf{c}$ to $\textbf{c}'$.

The Markov approximation based service placement policy algorithm is described in Algorithm 2, which can be implemented in the network operator that can gather sufficient network information and computing capability for real-time decision making. In the algorithm, a random service will be picked to update its placement policy for each update iteration. In this situation, a state transition of services from $\textbf{c}$ to $\textbf{c}'$ only occurs if only one user service is migrated. Since knowing the targeting migration policy performance (i.e., supposing we migrate service from MEC node $a$ to MEC node $c$, the new joint cost of perceived latency and migration in (\ref{P2}) can be calculated easily), the probability of each feasible migration adjustment is directly proportional to the difference of the total cost under two placement policies $\textbf{c}$ and $\textbf{c}'$, denoted as follows:
\begin{equation}
\label{transfer_prob}
	q_{\textbf{c},\textbf{c}'}(t) = \alpha \exp\Big(-\frac{1}{2}\beta\big(U(\textit{\textbf{c}$'$,{t}}) - U(\textit{\textbf{c},{t}})\big)\Big).
\end{equation}
Note that during each policy iteration, the network operator will record the best policy found up to now. As shown in \cite{Chen2013}, by proper parameter tuning Markov approximation algorithm can converge in a super-linear rate. Next, we analyze the complexity of the Markov approximation algorithm. For each update iteration, the system chooses arbitrarily a mobile device to update its service placement. During the process of calculating the total cost $U(\textit{\textbf{c$'$},{t}})$ for all feasible placement policies, the possible placement configurations enumerates at most $MN$. Assuming that this algorithm needs to be executed $I$ iterations to achieve the convergence, then the total time complexity of Algorithm 2 is $O(IMN)$. 
\subsection{Best Response Update Method}
To dramatically reduce running time of placement decision-making, we apply the best response update technique to construct a distributed mechanism for a faster service placement search. Different from the centralized method based on Markov approximation where the near-optimal placement decision is achieved by centralized probabilistic policy explorations collectively by all the users, in distributed service policy update, each user $k$ is generally greedy and adopts the best response to optimize its own placement decision in a deterministic manner. That is, best response update method emphasizes more on the exploitation of individual efficient decision instead of randomized decision exploration, leading to significant running time reduction.

As aforementioned description about edge resource allocation $D^k(t)$, the resource competition among multiple users will influence the service performance. Inspired by the application of game theory to non-cooperative AP channel selection \cite{xu13},
we consider the problem $\pmb{\mathcal{P}2}$ as a congestion game \cite{milchtaich1996congestion} with user-specific cost function. Let $\textbf{c}_{-k} = \{c_1,..., c_{k-1}, c_{k+1},..., c_N\}$ be service placement decisions made by all users except for user $k$. Given the placement policies of all other users $\textbf{c}_{-k}$ , the placement problem confronted by user $k$ is to select a proper MEC node to minimize its cost in terms of perceived latency and migration cost, i.e.,  
\begin{equation}
c_k = arg \min_{c_k \in {\mathcal{M}}}U_k(c_k, \textbf{c}_{-k}, t), \forall k \in \mathcal{N}.
\end{equation}
The non-cooperative nature of the service placement problem leads to a formulation based on game theory, where each placement decision is finally executed by the user device in a mutually acceptable way, i.e., a \textit{Nash equilibrium}, which is defined as follows:
\begin{myDef}
	(\textbf{Nash equilibrium.}) A placement policy profile $\textbf{c}^* = (c^*_1,..., c^*_N)$ achieves a Nash equilibrium when no user can minimize its cost further by unilaterally updating its placement policy, i.e.,
	\begin{equation}
	\label{Nash}
	U_k(c^*_k, \textbf{c}^*_{-k}, t) \le U_k(c_k, \textbf{c}^*_{-k}, t), \forall k \in \mathcal{N},\forall \textbf{c}_k \in {\mathcal{M}}.
	\end{equation}
\end{myDef}
To study the existence of the Nash equilibrium of multiple service placements, we introduce the best response update first.
\begin{myDef}
	\label{BRU}
	(\textbf{Best response update.}) Given the service placement profile $\textbf{c}^*_{-k}$ for all other users, the placement decision of user $k$ is the best response if 
		\begin{equation}
		U_k(c^*_k, \textbf{c}_{-k}, t) \le U_k(c_k, \textbf{c}_{-k},t), \forall c_k \in {\mathcal{M}}.
		\end{equation}
\end{myDef}
Similar to the n-player congestion game in \cite{milchtaich1996congestion}, through a finite best response update execution for service migration, our distributed service placement policy search can reach a Nash equilibrium by induction, i.e.,
Suppose that once a user sends the service request to the connected MEC node, the system will allocate a unique ID for its service. Then, we can update all service placement profiles according to the random order of assigned IDs. Let a service $k$, which is the current smallest index among the pending update users, be assigned to a preferable MEC node to achieve its current performance cost minimization by the best response update. Hence we can formulate the service placement update process as follows: 
\begin{eqnarray}
\begin{aligned}
\label{nonUpdate}
c_k(r+1) = arg \min U_k\big({c_k},\{{c}_1(r+1),..., {c}_{k-1}(r+1),\\
{c}_{k+1}(r),...,{c}_N(r)\},{t}\big),
\end{aligned} 
\end{eqnarray} 
where $r$ is the policy update round. In the current service placement profile, the services with smaller indexes have updated (i.e., ${{c}_1(r+1), ..., {c}_{k - 1}(r+1)}$), while the strategies of the ones with larger indexes are kept unchanged. By adopting the asynchronous best response update strategy, the service migration profile will be gradually converged. The detailed implementation is summarized in the Algorithm 3.

\begin{theorem}    
	\label{NashT}
	There exists a Nash equilibrium of the distributed service placement that can be achieved within at most $M\tbinom{N+1}{2}$ best response update steps.
\end{theorem}
The detailed proof is given in the technique report \cite{proof}.
Note that the proposed best response update based distributed policy search approach explores the possible service improvement update paths and terminates when achieving to a Nash equilibrium. Due to the weakly acyclic property (i.e., there must exist a finite improvement update path) \cite{milchtaich1996congestion}, it can make our policy coverage into a Nash equilibrium.  
\begin{algorithm}
	\caption{Best Response Update based Placement Policy Search} 
	\begin{algorithmic}[1]
		\State \textbf{Initialization:} Initialize the service placement profile $\textbf{c}(0) = ({c}_{1}(1),{c}_{2}(0),...,{c}_{N}(0))$ as randomly assigning a MEC node for each service and the update iteration round as $r=0$. 
		\State \textbf{End initialization}	
		\While {\textbf{c}(r) does not reach a Nash equilibrium}
		\For {indexed service $k$ = 1 to $N$}
		\State \textbf{Select} the proper MEC node where user $k$ can minimizes the its own cost acc. to (\ref{nonUpdate}) and gain the corresponding placement policy ${c}_k$
		\EndFor
		\State \textbf{Set} service migration profile as $\textbf{c}(r+1) = ({c}_1(r+1),...,{c}_N(r+1))$ and the update iteration round $r = r + 1$
		\EndWhile                                    
	\end{algorithmic}    
\end{algorithm}

Next, we evaluate the computational complexity of the algorithm 3. As shown in line 4 to 6, for each update iteration, the system will update service profile of users placement, which involves N minimization operations and each minimization operation can be achieved by sorting over at most M values. Hence this procedure has the complexity of $O(NM\log M)$. Line 7 has a complexity of O(1). Assuming that the algorithm 3 needs $I$ times update iteration to be converged to a Nash equilibrium. Then the total computational complexity of the algorithm 3 is $O(INM\log M)$. Surprisingly, the total computational complexity of distributed placement update seems to be higher than Markov approximation. While, in practical process, the update iteration round is a critical factor to increase the time overhead. In the later simulate experiment, we can find that the distributed scheme reduces dramatically running time of placement decision-making compared with the Markov approximation.

\section{Performance Analysis}
\label{verification}

In this section, we analyze theoretically the performance of our two {mobility-aware dynamic service placement algorithms for MEC}. First, we discuss the optimality gap in the Markov approximation based scheme and best response update based scheme respectively. Then, we compare the performance of our two online algorithms (i.e., Markov approximation and best response update in the Lyapunov framework) with the offline optimum.

\subsection{Markov Approximation}
With above description of Markov approximation algorithm, the probability of a service placement state switch from $\textbf{c}$ to $\textbf{c}'$ in the Markov chain is denoted in (\ref{transfer_prob}). It is obvious that our algorithm can be converged to a distinctive stationary distribution for its time reversibility.

\begin{theorem}    \label{Markov}
	There exists a distinctive stationary distribution for the service placement algorithm as stated in equation (\ref{stationaryDistribution}).
\end{theorem}

The detailed proof is given in the technique report \cite{proof}.
As shown in Theorem 1, we can obtain the minimal supremum bound for the \emph{drift-plus-penalty} function as the parameter $\beta$ increasing to a large enough value in our service placement algorithm. We denote the minimal supremum bound and expected supremum bound by proposed algorithm as $S^*= \min\sum_{\textbf{c}\in c(t)}U(\textit{\textbf{c},{t}})$ and $\widetilde{S} = \sum_{\textbf{c}\in c(t)}q^*_\textbf{c}U(\textit{\textbf{c},{t}})$ respectively.

\begin{theorem}    \label{long_sum}
	For the algorithm, the optimality gap is given as follows:
		\begin{eqnarray}
		0 \le  \widetilde{S} - S^* \le \frac{1}{\beta}\ln|\delta|,
		\end{eqnarray}
	where $|\delta|$ is the amount of feasible service placement policies of all mobile users at time slot $t$.
\end{theorem}

The detailed proof is given in the technique report \cite{proof}. By Theorem 2, the error of worse-case solution in our algorithm is no more than $\frac{1}{\beta}\ln|\delta|$. Thus, if setting the value of parameter $\beta$ as large as possible, we can approach an almost equivalent solution to the minimal supremum bound. Fortunately, the value of $\beta$ is usually large enough in an acceptable scope, the performance deviation of the optimum is quite small \cite{Chen2013}. 

\subsection{Approximation Ratio for Best Response Update}

With above description of best response update, we can quantify the efficiency ratio of our distributed placement mechanism in the worst-case equilibrium over the optimal centralized one. Let $\Gamma$ be the set of equilibria of the service placement profile. Then the approximation ratio for best response update can be expressed as follows:

	\begin{equation}
		\mu = \frac{\max_{\textbf{c} \in \Gamma }\sum_{k \in \mathcal{N}} U_k(c,t)}{\min_{\textbf{c} \in c(t)} \sum_{k \in \mathcal{N}} U_k(c,t)},
	\end{equation}

Obviously, the lower bound of approximation ratio $\mu$ is 1. A larger approximation ratio denotes that the worst performance of our distributed algorithm is less efficient than using the centralized optimum as a benchmark. Let $F_{max}$ and $F_{min}$ be the maximum and minimum computing capacity of all MEC nodes respectively. Similarly, Let $H_{max}^k = \max_{i \in \mathcal{M}, t\in \mathcal{T}} H_i^k(t)$, $H_{min} = \min_{i \in \mathcal{M}, t\in \mathcal{T}} H_i^k(t)$, $R^k_{max} = \max_{t\in \mathcal{T}} R^k(t)$, $R^k_{min} = \min_{t\in \mathcal{T}} R^k(t)$ and $\rho_{max} = \max_{i \in \mathcal{M}, t\in \mathcal{T}} \rho_i^k(t)$. Thus, we can have:

\begin{lemma}
	\label{Nash}
	In the distributed service placement search, the joint cost performance of each user $k$ at an equilibrium is no more than $\frac{VR^k_{max}(M+N-1)}{MF_{min}} + VH_{max}^k + \rho_{max}^k$.
\end{lemma}

The detailed proof is given in the technique report \cite{proof}. According to Lemma 2, we can gain the upper bound of the approximation ratio $\mu$ as follows:

\begin{theorem}
	\label{PoA}
	The the approximation ratio $\mu$ of the distributed service placement search is at most
	\begin{small}
	\begin{equation}
		\mu \le \frac{\frac{VR^k_{max}(M+N-1)}{MF_{min}} + VH_{max}^k + \rho_{max}^k}{\sum_{k=1}^{N}\big(\frac{VR_{min}^k}{F_{max}} + VH_{min}^k \big)}.
	\end{equation}
	\end{small}
\end{theorem}
The approximation ratio $\mu$ demonstrates the worse-case performance of our distributed scheme in an equilibrium. Numerical results in the next section show the algorithm is efficient compared with the centralized approximation. 
\subsection{Optimality Analysis} 
As we have mentioned, the transformed problem $\pmb{\mathcal{P}2}$ is NP-hard. Fortunately, the minimization error of the $\pmb{\mathcal{P}2}$ is acceptable under the control of our online algorithm. We use $\widetilde T^k(t)$ and $T^{opt}$ to respectively denote the delay performance in time slot $t$ by the proposed algorithm and the infimum time average performance delay with the overall information. Then the following theorems will give a supremum bound of the time-averaged delay performance and the migration cost queue backlogs for our two approximation algorithms. The former one is the Markov approximation based centralized scheme, and the later one is the best response update based distributed scheme.

\begin{theorem}    \label{performance-1}
For any non-negative control parameter $V$, the long-term delay performance implemented by proposed two online algorithms satisfy that
\begin{eqnarray}
\label{theorem3}
\lim_{T \to \infty} \frac{1}{T}\sum_{t = 0}^{T-1}\sum_{k = 1}^{N}\mathbb{E}\{\widetilde T^k(t)\} \le T^{opt} +\frac{B}{V} + \frac{1}{\beta V}\ln|\delta|.
\end{eqnarray}
\vspace{-15pt}
\begin{eqnarray}
\label{theorem3}
\lim_{T \to \infty} \frac{1}{T}\sum_{t = 0}^{T-1}\sum_{k = 1}^{N}\mathbb{E}\{\widetilde T^k(t)\} \le \mu T^{opt} +\frac{B}{V}
\end{eqnarray}
\end{theorem}
\begin{theorem}    \label{performance-OCA}
Assuming that $E_{avg} > 0$ and initializing the migration cost queue backlog is 0, thus for all time slots we having the following bound: 
\begin{eqnarray}
\label{theorem4}
\lim_{T \to \infty} \frac{1}{T}\sum_{t = 0}^{T-1}\mathbb{E}\{Q(t)\} \le \frac{B + VT^{opt}}{\varepsilon} + \frac{1}{\beta \varepsilon}\ln|\delta|.
\end{eqnarray}
\vspace{-15pt}
\begin{eqnarray}
\label{theorem4}
\lim_{T \to \infty} \frac{1}{T}\sum_{t = 0}^{T-1}\mathbb{E}\{Q(t)\} \le \frac{B + \mu VT^{opt}}{\mu\varepsilon}.
\end{eqnarray}
\end{theorem}

The detailed proof is given in the technique report \cite{proof}. 
Where $\varepsilon > 0$ is a finite constant that represents the distance between the time-averaged migration cost by some control policy and long-term cost budget. From Theorem 1, it is known that the delay performance of the online algorithm can be approached closely to the offline optimum with the adjustable control parameter increasing $V$. Besides, the bound of migration cost queue backlog is also determined by the parameter V. In short, a performance-cost trade-off of $[O(1/V),O(V)]$ exists in our online algorithm, where we can set the parameter $V$ to a desirable value to achieve the balance of the long-term delay performance and migration cost.

\section{Performance Evaluation}
\label{OCA}

In this section, we conduct numerical studies to evaluate the time-averaged perceived latency performance under the long-term migration cost constraint of the proposed algorithms and to verify the derived theoretical results.
\begin{figure*}[htbp]
	\centering
	\begin{minipage}[t]{0.5\linewidth}
		\subfigure[Average perceived latency performance with different values of control parameter $V$ under different placement policies] {\includegraphics[width=2.8in, height = 2.3in]{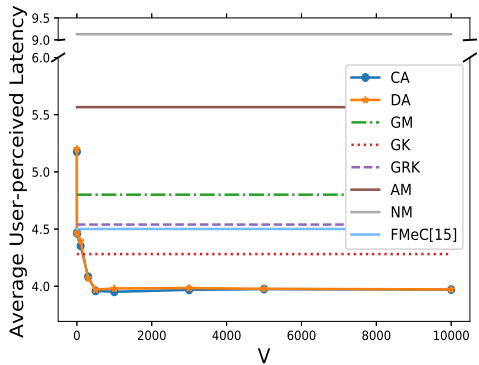}}
	\end{minipage}
	\begin{minipage}[t]{0.4\linewidth}
		\subfigure[Average perceived latency performance with different long-term cost budgets $E_{avg}$] 
		{\includegraphics[width=2.8in]{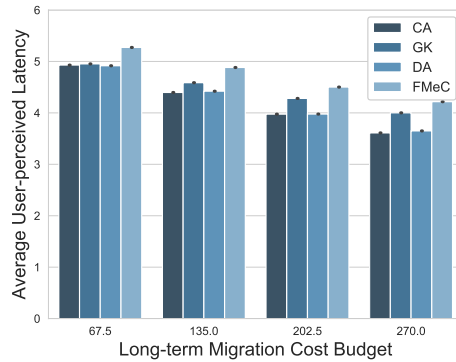}}
	\end{minipage}
	\caption{ Optimality analysis }
	\label{fig2}
\end{figure*}
\subsection{Simulation Setup}
We adopt the ONE simulator \cite{Keranen2009} to conduct system simulation, where mobile devices move along the roads or streets based on Shortest Path Map-Based Movement Model \cite{Keranen2009} in a downtown Helsinki, Finland. For simplicity, we divide the whole area into 63 square parts. Each part occupies 500$\times$500 $m^2$, endowed with one MEC node to provide mobile services. Each MEC sever is equipped with multiple CPU cores, and the maximum computing capacity $F_i = 25GHz$. Besides, considering diverse mobility patterns in realistic environment, we choose two typical kinds of mobile users: about $85.7\%$ $(\frac{6}{7})$ of the mobile users is pedestrians with speed uniformly distributed in $[0.5, 1.5]$ m/s, the remaining users are drivers with speed uniformly distributed in $[2.7, 11.1]$ m/s. The hop distance between two MEC nodes is calculated by Manhattan distance. We simulate 2000 time slots for our system, and the interval of a time slot is 5 minutes. During each time slot $t$, we assume that the placement for service profile of users and wireless connections between user and edge are unchanged. The request arrival process in the interval $R^k(t)$ for each user $k$ is uniformly distributed within $[0.6, 1]$ Mbps and its processing density is 2640 cycle/bit (such as 400 frame video game in \cite{kwak2015dream}). To simplify the problem, we assume that the maximum computing capacity of all MEC nodes is the same and the current communication delay follows uniform distribution within $[1, 1.35]$ of the optimal delay, which is 0.6 min per hop for every service. It is the same as migration cost, perturbed by timing a random parameter in $[1, 1.35]$. The difference is that one hop migration takes 1 unit cost, and plus 0.5 unit cost in the end, which is allied to the request arrival process.

\subsection{Performance Benchmark}
We consider two representative situations and four approaches as a benchmark to evaluate our algorithms. One situation is no matter what the distribution of mobile user is, the service VM is always migrated to execute on its nearest MEC node, i.e., "Always migration" (AM) strategy. On the contrary, another is always to keep the initial assignment policy unchanged ("No migration" ) strategy.
Furthermore, the four algorithms are described as follows:
\begin{itemize}
	\item[1)] 
	\emph{GM}: this algorithm migrates the request services to the nearest MEC node at every opportunity over a long period of time. 
	\item[2)] 
	\emph{GRK}: this algorithm randomly picks up different $K$ services and migrates them to the current optimal MEC nodes at every opportunity over a long run.
	\item[3)] 
	\emph{GK}: this algorithm migrates $K$ services in descending order by time cost to the current optimal MEC nodes at every opportunity over a long run.
	\item[4)] \emph{FMeC} \cite{aissioui2018enabling}: this algorithm migrates services to the current optimal MEC node based on the mobility pattern update analysis at every opportunity over a long run. Since the one-dimensional (1D) mobility model with one direction of traffic flow was assumed in \cite{aissioui2018enabling}, we assume the estimated direction of velocity in next time slot is similar to the current time slot in our simulation.
\end{itemize} 

\begin{figure}
	\vspace{-10pt}
	\centering
	\subfigure[Average migration cost queue with different values of control parameter $V$]{
		\includegraphics[width=2.8in]{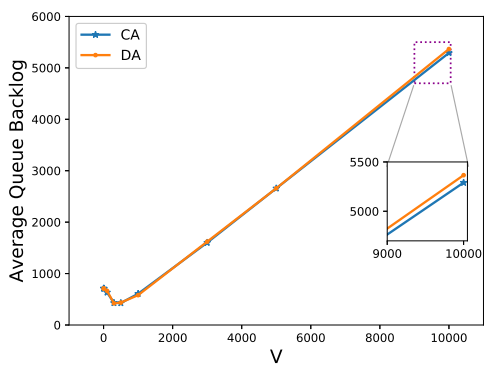} 
	}  
	\subfigure[Average migration cost queue with different values of control parameter $V$ at different time slots]{
		\includegraphics[width=2.8in]{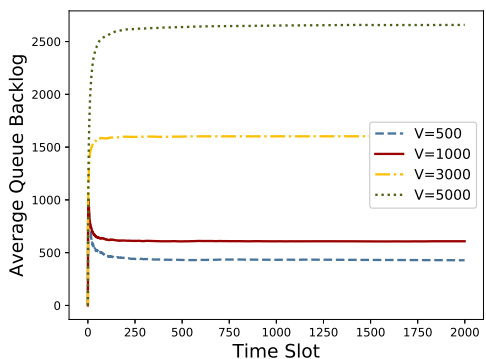} 
	}  
	\caption{Queue stability}
	\vspace{-15pt}
	\label{fig:my_label}
	\hfill
\end{figure}
\subsection{Latency Cost Trade-off}
Obviously, the optimization of the user-perceived latency and migration trade-off in a cost-efficient manner is the key to the long-term service placement problem, which guides the following analysis for our proposed two algorithms: Markov approximation based centralized algorithm (CA) and best response update based distributed algorithm (DA). 

\emph{Average user-perceived latency optimality.} To analyze key elements to influence the user-perceived latency, we formulate a standard of comparison, where 315 mobile users move in the city and the long-term time-averaged migration cost budget for the network operator is set as 202.5 cost units, approximation control parameter $\beta$ is set as $0.1$.  
Fig. 2(a) shows the average latency with different values of control parameter $V$ under various online algorithms. We can observe that the average latency decreases with $V$ increasing, and gradually approaches a minimum value in both two proposed algorithms (i.e., CA and DA). This confirms Theorem 5 we have mentioned in the theoretical analysis that the time-averaged latency performance is proportional to the $1/V$. Besides, compared with the benchmark, our two algorithms do have remarkable improvements in average latency performance, around from 8\% to 56\% improvement with $V = 1000$. For the AM strategy, the major reason for the poor performance is the low utilization of edge resources. In general, only almost two-thirds of the MEC nodes provide all user services during every time slot $t$. Even though GRK and GK make up this deficiency of the inefficient utilization, the unreasonable migration policy still exists since every migration selection update is a local optimization. Furthermore, we find the particular migration sequence, such as descending order, can alleviate the performance gap to some extent. The overload among MEC nodes and inaccurate mobility pattern analysis deteriorate the latency performance of the FMeC algorithm. 
Intuitively, a larger migration cost budget can provide more opportunities for further enhancements on the placement optimization. As illustrated in Fig. 2(b), with the cost budget increasing, CA and DA have more notable improvements compared with GK and FMeC algorithms. 

\begin{figure}
	\vspace{-15pt}
	\centering
	\subfigure[Average migration cost with different values of control parameter $V$]{
		\includegraphics[width=2.8in, height = 2.3in]{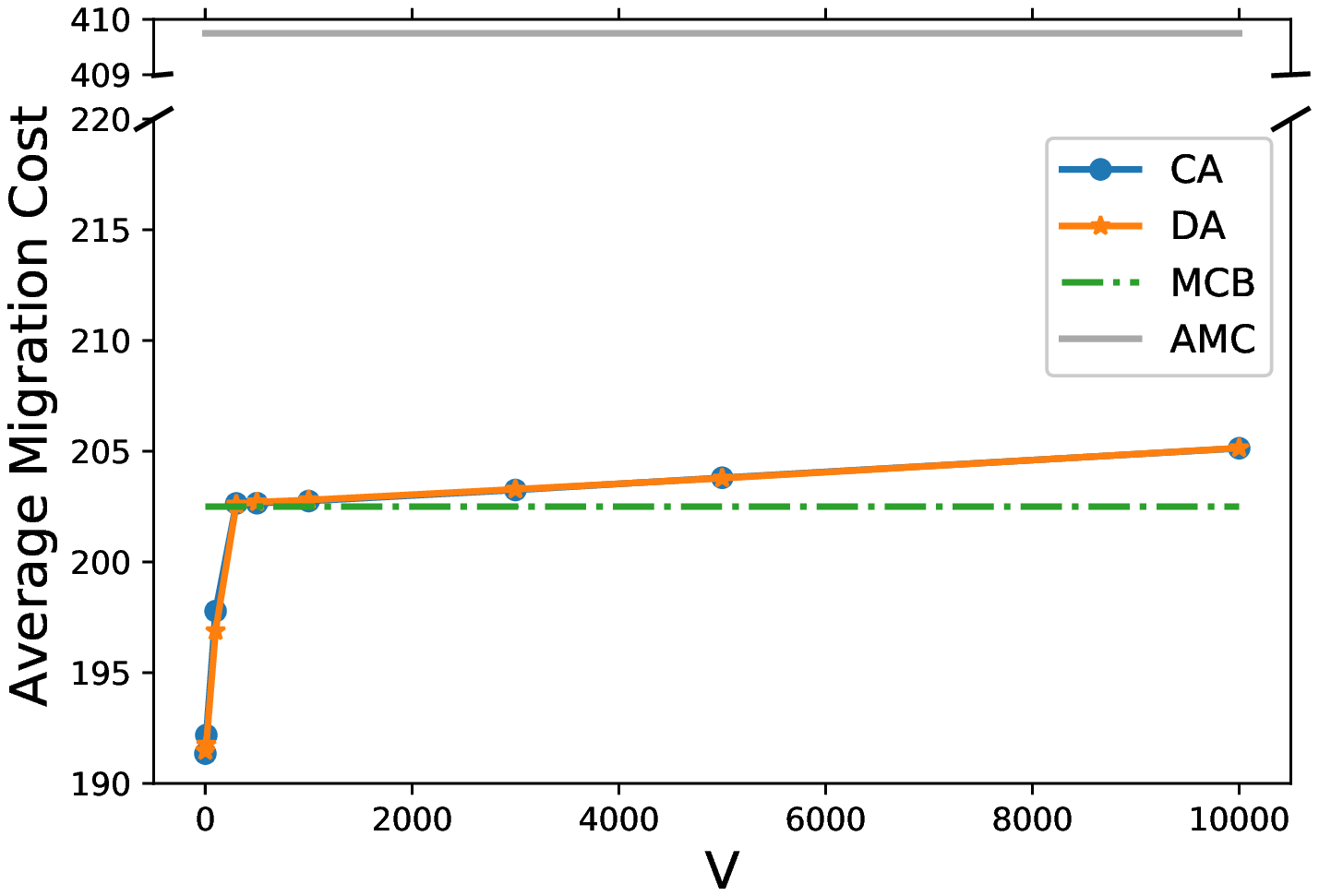} 
	}  
	\subfigure[Average migration cost with different values of control parameter $V$ at different time slots]{
		\includegraphics[width=2.8in]{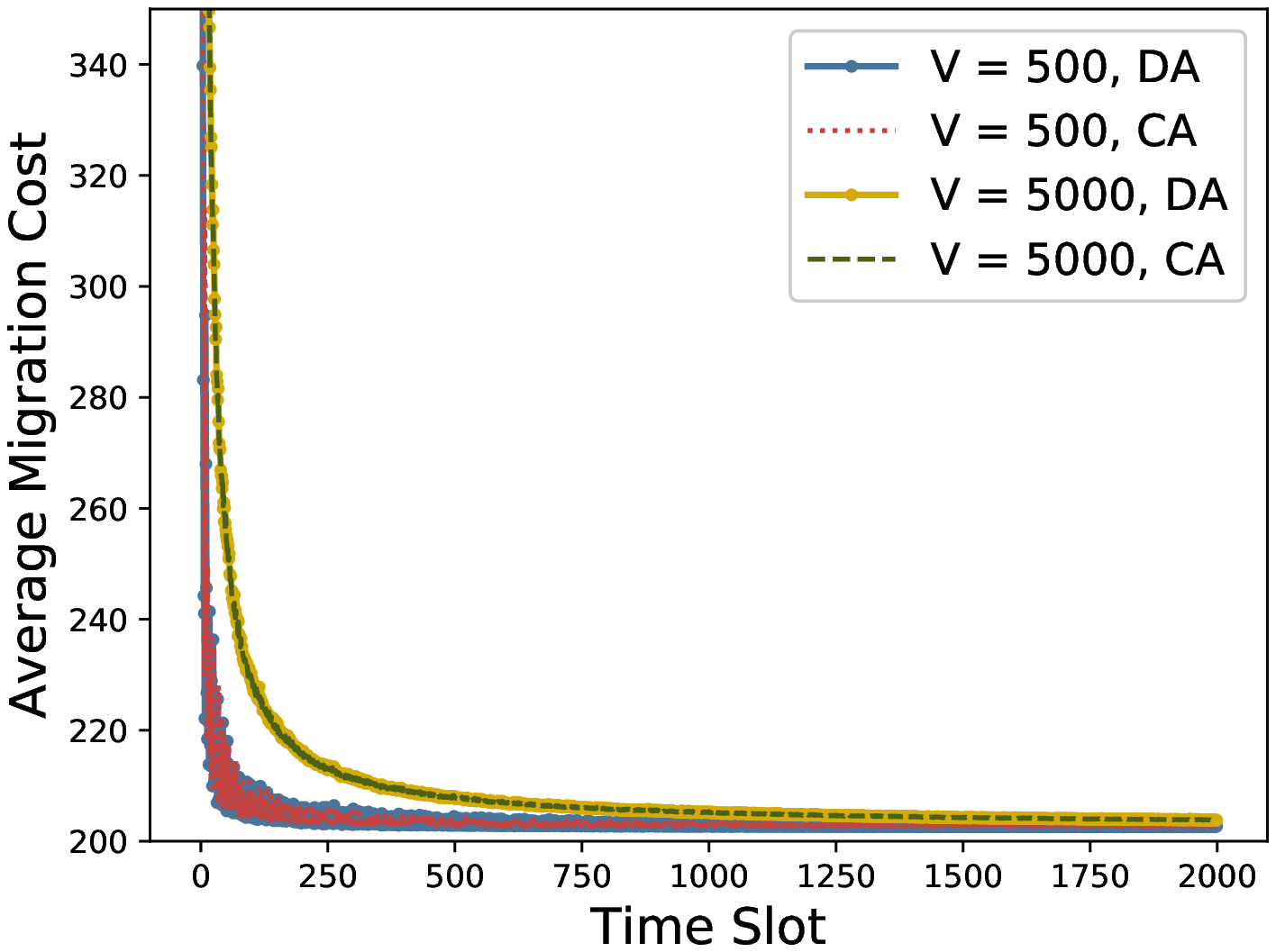} 
	}  
	 \caption{Convergence of average migration cost}
	\label{fig:my_label}
	\vspace{-15pt}
\end{figure}

\emph{Queue stability.} Fig. 3 (a) compares the time-averaged migration cost queue between CA and DA algorithms with different values of control parameter $V$. Broadly, as $V$ increases, the time-averaged backlog queue increases in a linear fashion, which is matched in Theorem 6. Besides, the CA scheme has a slightly better performance in queue backlog with a large value of $V$. 
Along with Fig. 2(a), the performance of time-averaged latency and migration cost follows the [$O(1/V),O(V)$] trade-off. As shown in Fig. 3(b), the varying curve of average migration backlog queue gradually becomes stable in our algorithm no matter what V is. It implies our proposed algorithms will satisfy the long-term cost budget, and the detailed discussion is presented later.  

\emph{Convergence of average migration cost.}
Fig. 4(a) plots the average migration cost with different values of $V$ under two proposed algorithms. Note that the migration cost budget (MCB) is almost half of the all services migration cost (AMC). In this situation, a large value of $V$ makes system care more about user-perceived latency, which may violate the long-term cost budget in finite time slots, such as $V = 5000$. While in Fig. 4(b), as time slot increases, the average migration cost decreases remarkably and gradually converges to the migration cost budget under different values of control parameter $V$. The reason for this problem is insufficient time slots. As we have discussed, if the migration cost queue is stable, i.e., $\lim_{T \to \infty}  \mathbb{E}\{Q(T)\}/T=0 $, the actual migration cost would not violate the budget. In Fig. 3(b), we know that all migration backlog queues gradually converge to some certain finite value. Thus, if increasing time slots, the long-term constraint can be satisfied. 

\begin{figure}

	\subfigure[Dynamic adjustment for migration cost queue in Markov approximation based scheme]{
		\includegraphics[width=2.8in]{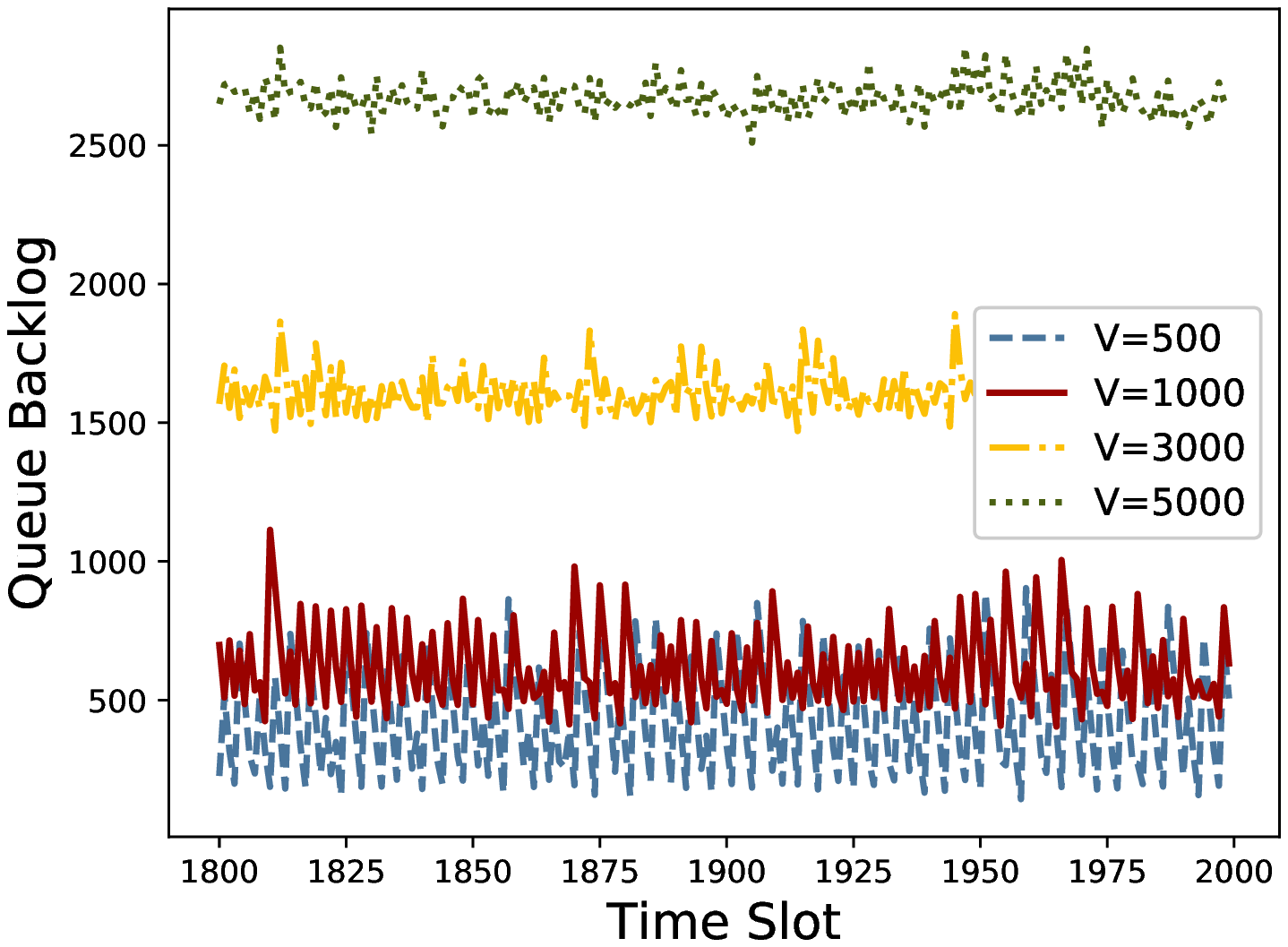} 
		}  
		\subfigure[Distribution of user-perceived latency with different values of control parameter $V$]{
			\includegraphics[width=2.8in]{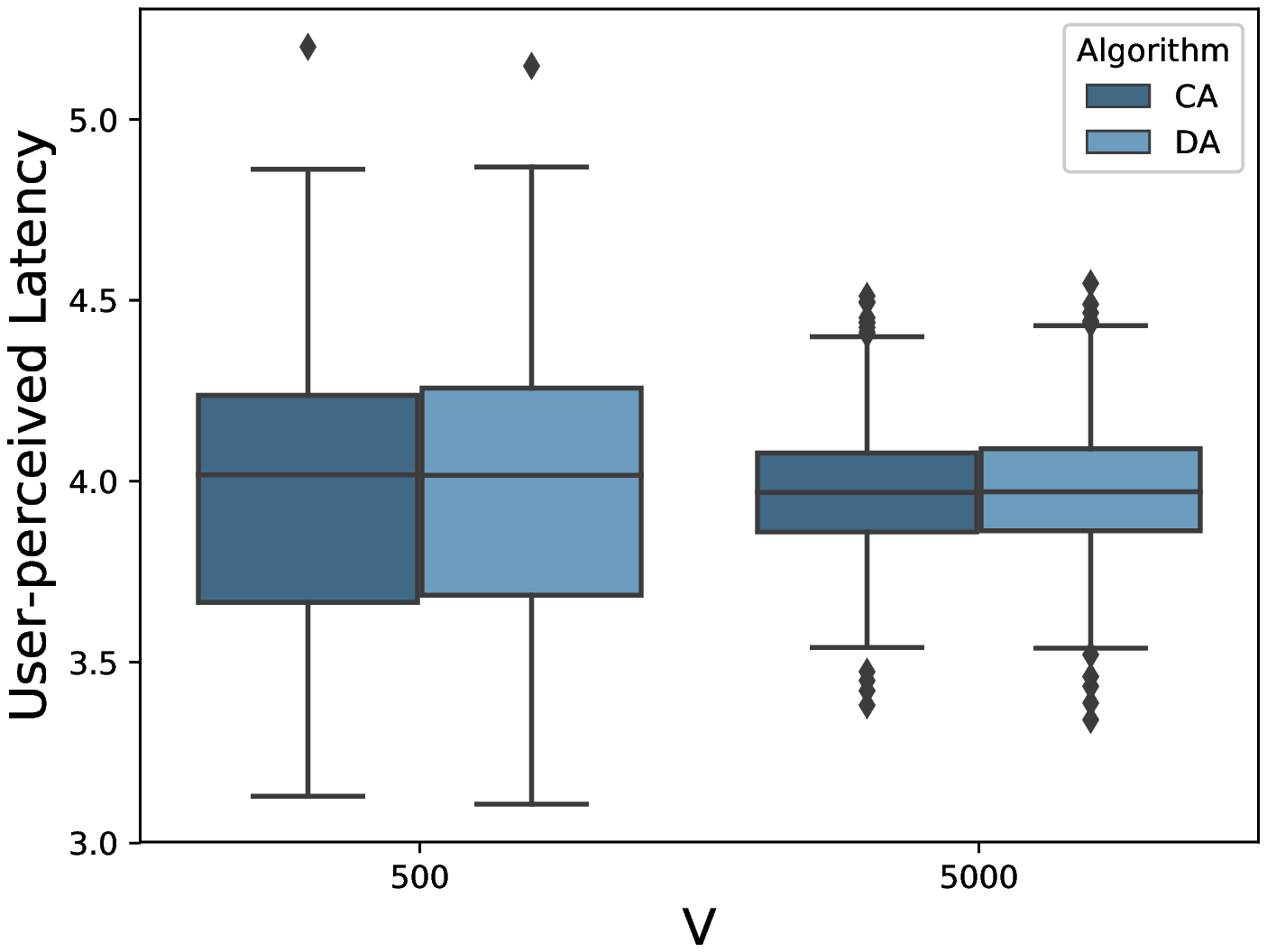} 
		}  
		\caption{Adaptability to system dynamics}
		\label{fig:my_label}
	\vspace{-15pt}
	\hfill
\end{figure}

\emph{Adaptability to system dynamics.} To further explore the dynamic adjustment of our algorithm to minimize the average latency performance under the control of migration cost budget, we depict a part of real-time fluctuation of the migration backlog queue and the distribution of latency performance with different values of $V$. As shown in Fig. 5(a), for a clear fluctuation exposition, we select a partial time snippet from the long run. It can be observed that the real-time migration backlog queue fluctuates frequently, which means the system will adjust migration policy frequently. When the current migration backlog queue is large and thus the remaining available migration cost is relatively scarce, the CA algorithm will endeavor to reduce the queue backlog to prevent the over-budget. Contrarily, when the current migration backlog queue is small, minimizing the time cost is the prime goal since the remaining available migration cost is abundant. Surprisingly, the large value of $V$ reduces the fluctuation range of the migration backlog queue, which makes system real-time latency performance more stable. Fig. 5(b) compares the distribution latency performance between two algorithms with different values of $V$. It is obvious that the distribution of latency performance is more centralized to the median as the control parameter $V$ increasing, which is consistent with the dynamic adjustment of migration backlog queue.

\subsection{Efficient on different dense networks}
\begin{figure}
	\centering
	\includegraphics[width=2.8in]{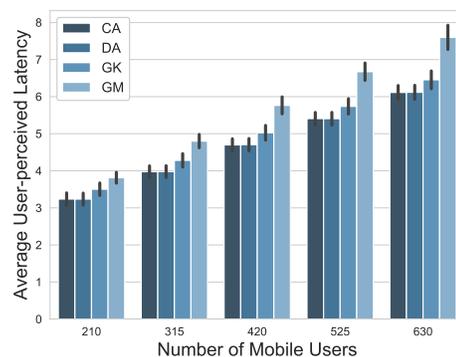} 
	\caption{Average perceived latency performance in different dense networks}  
	\vspace{-10pt}
\end{figure}
Fig. 6 suggests that our two algorithms still work efficiently in different dense networks (i.e., the percentage ratio of user amount to MEC node amount). Higher user dense network will lead to the rapid growth of computation delay, which is the main factor of perceived latency increasing. Our algorithm can balance the edge load by migrating services to slow the total perceived latency growth. Nevertheless, the computation delay has been growing significantly faster than total perceived latency, while the network propagation delay is not affected by user amount. To maintain the original quality of service, the network operator should improve the computation capacity of edges accordingly.

\subsection{Process Time for Placement Decision-Making}
We evaluate the proposed algorithms on Intel Core i7-6700 CPU (4 Cores @ 3.4G) computer with PyCharm. Fig. 7 suggests that the distributed scheme (DA) can dramatically reduce the running time of placement decision-making compared with the Markov approximation based scheme (CA), especially when the amount of users is large. The critical reason is that the asynchronous best response update can converge faster into an equilibrium. This result demonstrates the distributed scheme is more efficient in a large-scale and ultra dense network. Besides, the trend of curves for both two algorithms is consistent with their total time complexities.  
\begin{figure}
	\centering
	\includegraphics[width=2.8in]{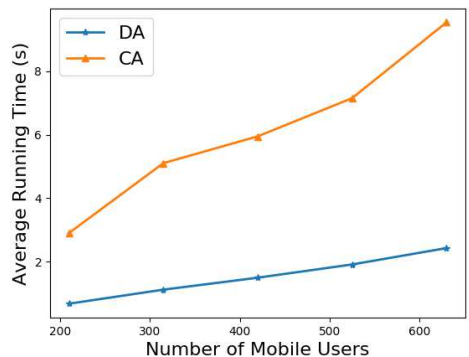} 
	\caption{The average running time for placement update in different dense networks}
	\vspace{-10pt}
\end{figure}

\section{Conclusion}
\label{sec:con}
In this paper, we study the mobile edge service performance optimization problem with long-term time-averaged migration cost budget. We design a novel mobility-aware online service placement framework to achieve a desirable balance between time-averaged user-perceived latency and migration cost. To tackle the unavailable future system information, which involves mobility pattern and request arrival processes, we utilize Lyapunov optimization technique to incorporate the long-term budget into a series of real-time optimization problems. Since the decomposed optimization is an NP-hard problem, we develop two efficient heuristic schemes based on the Markov approximation and best response update techniques to approach a near-optimal solution. Furthermore, we provide the theoretical proof of our proposed algorithm performance. Besides, extensive simulation demonstrates the effectiveness of our online algorithm while maintaining the long-term migration cost constraint. Future research direction includes combining online network selection in ultra dense networks. To gain a lower end-to-end latency, a user will choose an optimal one (such as a less congestion one) out of multiple access networks (e.g., cellular macrocell, femtocell, and WiFi networks). Further, we aim to extend the existing federated edge cloud framework by incorporating the Device-to-Device (D2D) collaboration, where mobile users can beneficially share their computation and communication resource in a closer proximity.


\bibliographystyle{IEEEtran}
\bibliography{ref}
\end{document}